# Fast-light Assisted Four-Wave-Mixing in Photonic Bandgap

Cheng Feng[1], Hao Luo[1], Liang Zhang[1], Jinmei Liu[1,2], Li Zhan[1,*]

Abstract: Since the forward and backward waves are coupled with each other and a standing wave with no net propagation of energy is formed in the photonic bandgap, it is a commonsense of basic physics that, any kinds of effects associated with wave propagation including four-wave-mixing (FWM) are thought to be impossible. However, we lay great emphasis here on explaining that this commonsense could be broken under specific circumstances. In this article, we report with the first experimental observation of the energy conversion in the photonic bandgap into other channel via FWM. Owing to the phase manipulation by fast light effect in the photonic bandgap, we manage to achieve the phase-match condition and thus occurred FWM transfer energy into other channels outside the photonic bandgap efficiently. As one-dimensional photonic crystal, simulations on fiber Bragg grating (FBG) with and without fast light were conducted respectively, and an enhanced FWM in photonic bandgap of FBG was observed. The experimental result shows great agreement with the analysis.

[1]Department of Physics and Astronomy, Key Laboratory for Laser Plasmas (Ministry of Education), State Key Lab of Advanced Optical Communication Systems and Networks, Shanghai Jiao Tong University, Shanghai, 200240, China. [2]Department of Physics, East China Normal University, Shanghai, 200240, China. Correspondence and requests for materials should be addressed to Li Zhan. (e-mail: lizhan@sjtu.edu.cn)

Photonic bandgap can be created in various kinds of periodic photonic media, in which photonic energy cannot propagate through the medium[1,2]. Due to operation in the photonic forbidden band, any kind of phenomena related with light propagation including nonlinear effects such as the four wave mixing (FWM) is thought to be impossible. Furthermore, since forward and backward waves are coupled with each other and a standing wave is formed, net propagation of energy is also forbidden. In order to convert energy out of the photonic bandgap, reports about the transmissivity enhancement in photonic bandgap by inducing an extra phase during nonlinear process[3,4], especially self-phase modulation (SPM) arouses great interest and an enhancement of 20dB has been achieved on researches in Bragg soliton[5]. However, the high power (~1kW) in these experiments remains this kind of energy conversion unfeasible[3–5].

In contrast, slow/fast light effect[6–11] offers a new way to control the speed of light as well as to manipulate optical phase in a rather low energy level. Since the 1st derivation of propagation coefficient is the reciprocal of the group velocity $v_g$, namely $\beta_1 = d\beta/d\omega = 1/v_g$, the change of group velocity by fast light in the photonic bandgap and slow light at the band edges lead to the variation of propagation coefficient $\Delta\beta$. Thus an extra phase along propagation $\Delta\varphi = \Delta\beta z$ is induced. By this way, optical phase can be manipulated, and the enhancement of spectral sensitivity of interferometer by a factor of maximum 100 has been reported[12–14]. FWM is a phase-match dependent optical nonlinear effect without threshold [9], so that phase-match can be satisfied by this introduced phase to enhance the FWM. Recently, experiments about controlling wavelength conversion efficiency via stimulated Brillouin scattering (SBS) slow light[15], and FWM process in slow light photonic crystal waveguide[16–18] or via slow light in electromagnetic induced transparency[19,20] has been reported. The FWM enhancement via other feasible slow/fast light methods, such as coherent population oscillation[8,10], stimulated Raman scattering[21], cross-gain modulation[22] and other structural slow/fast light methods remains undone. Conventionally, the slow-light effect is used to enhance optical nonlinear effect owing to the enhancement of power density in slow light process, but the fast light is not.

In our work, fast light effect in the photonic bandgap of fiber Bragg grating (FBG) is used to control the optical phase by inducing a large chromatic dispersion. Simulation as well as experiment in a special-designed FBG was carried out, and FWM process in the photonic bandgap has been observed without transmissivity increasing. This proves the FWM generation in photonic forbidden band by fast light effect rather than slow effect. With a rather low power (~200mW), our scheme involves little interference from unnecessary nonlinear effects.

It is well known that, FBG is a key component in many fields because of its excellent behavior as optical filter[23], and also it is a device of photonic bandgap. It exhibits strong dispersion both in reflection and transmission[24,25]. Thus fast light effect is obvious owing to anomalous dispersion in the bandgap[26]. The characteristic of a uniform FBG can be explained by the coupled-mode theory[27,28]. Its feature can be obtained by the amplitude reflection coefficient [29,30]. Because a large chromatic dispersion is induced near or in the photonic bandgap of FBG[31], the optical phase is supposed to vary rapidly near the bandgap[29]. Thus it causes a considerable group delay or advancement. However, the light transmission is forbidden in the bandgap. Since the group delay in transmission equals to the one in reflection[32], the group delay transmitted through a FBG is shown to be[33]

$$\tau(\lambda)_t = \tau(\lambda)_r = -\frac{\lambda^2}{2\pi c}\frac{d\varphi}{d\lambda} \qquad (1)$$

where $\varphi$ is the transmission phase. Due to the same phase shifts induced in two propagation directions, forward- and backward- waves remain coupled to each other[28]. Thus, even if slow/fast light effect is considered, the forward propagation in a FBG is still forbidden in the photonic bandgap.

**Results**

**Transmittivity and group delay of fiber Bragg grating.** As shown in Figure 1, we measured the transmission features of a Gaussian-apodized FBG with a length of 10cm, Bragg wavelength at 1550.186 nm, which was used in the following experiment. The optical phase as well as the group delay could also be derived. The maximum delay of 68.48ps at the band edge and the advancement of 68.71ps in the bandgap center was measured.

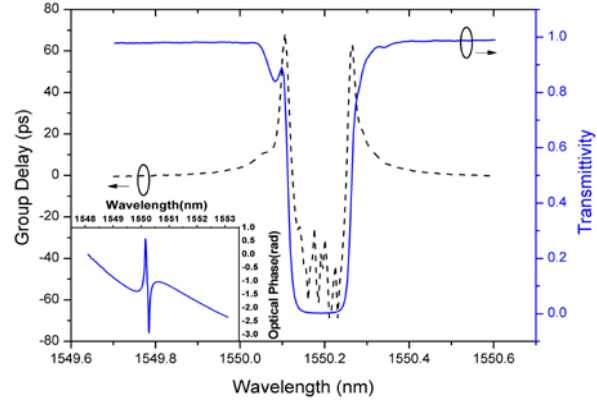

**Figure 1. Transmittivity and group delay of the FBG.** Group delay of 68.48ps at left, 62.75ps at right band edge and advancement of 68.71ps in the bandgap center were measured. Inset is the plot of optical phase as a function of wavelength. Due to the sensitivity limit of photodetector, the fluctuation of transmission time may occur in photonic bandgap.

Discontinuity in the optical phase results in slow/fast light effect near or in the bandgap. The optical phase passing the FBG is rewritten as

$$\varphi = \beta z \approx [\beta_0(\omega) + \beta_1(\omega)(\omega-\omega_0) + \frac{1}{2}\beta_2(\omega)(\omega-\omega_0)^2]z \quad (2)$$

where $\beta$ is the propagation coefficient expanded to second order in a Taylor series. When group velocity dispersion (GVD) $\beta_2 = d^2\beta/d\omega^2$ is considered, $\beta_1 = d\beta/d\omega = 1/v_g$ is no longer constant but increases at the edge of FBG and decreases in the bandgap. Therefore, with the variation of group velocity, the propagation coefficient and the optical phase can be manipulated.

**Simulation results of four wave mixing spectra with and without fast light.** Without considering the other nonlinear effects under low pump level, a degenerated FWM is controlled by the phase-match condition written as[34]

$$\Delta\beta = 2\beta_p - \beta_s - \beta_i \quad (3)$$

where $\beta_{p,s,i}$ are the propagation coefficients of the pump, signal, idler wave respectively. With the manipulation of the propagation coefficient via group velocity control, the phase-match condition $\Delta\beta = 0$ can be achieved. Considering the chromatic dispersion induced by fast light effect in the bandgap of FBG[35], the calculation shows, the dispersion of 3919.71ps/nm with the corresponding advancement of ~66.62ps can well compensates the phase-mismatch between ~0.5nm channel spacing in FWM experiment.

In order to verify the above analysis, simulations of degenerated FWM with a uniform FBG were conducted. Figure 2 shows the simulation results of FWM in FBG when fast light effect is not considered. With the pump wave approaching the photonic bandgap, only part of its energy is involved in the FWM process. Thus, the idler power together with pump power decreases magnificently. When the pump wavelength locates in the photonic bandgap, the idler waves in both sides are under noise level. In contrast, when the large chromatic dispersion induced by FBG and fast light effect is considered, a maximum power change of 7.14dB on the left side and 12.32dB on the right side is shown in Figure 3. Obvious decreases of idler power at 1549.14nm, 1549.28 nm on the left and 1550.47nm, 1550.88nm on the right correspond to the pump wavelength at the band edges, owing to the slow-light effect at these edges[29]. Meanwhile, a large enhancement of idler power matches well with pump wavelength in the photonic bandgap. Although the forward-propagation wave does not transmit to the end of FBG, the energy conversion can still be achieved by the satisfaction of phase-match condition with fast light effect as long as forward wave can propagate a short distance within the FBG[27]. Thus, even though the forward wave is forbidden on transmission, it still is involved in the FWM process.

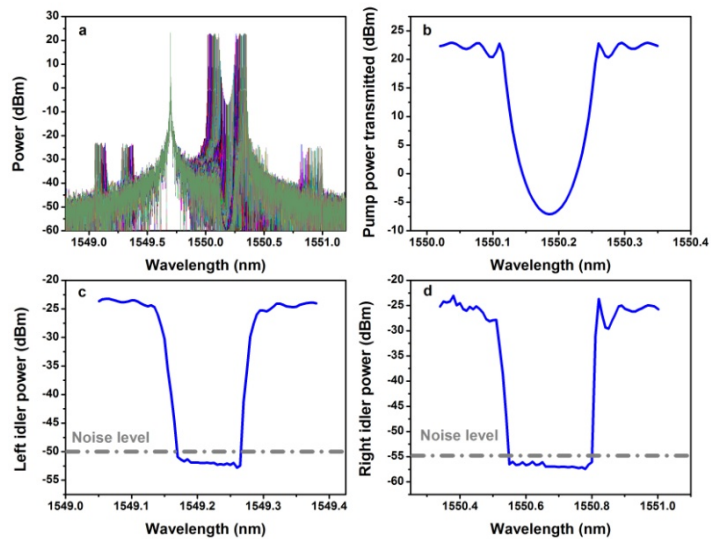

**Figure 2. Simulation result of degenerated FWM process without fast light**. Inset(a) shows the superposition of FWM spectrums with different pump wavelengths, and (b),(c),(d) shows the transmission spectrum of pump power, left and right idler spectra respectively.

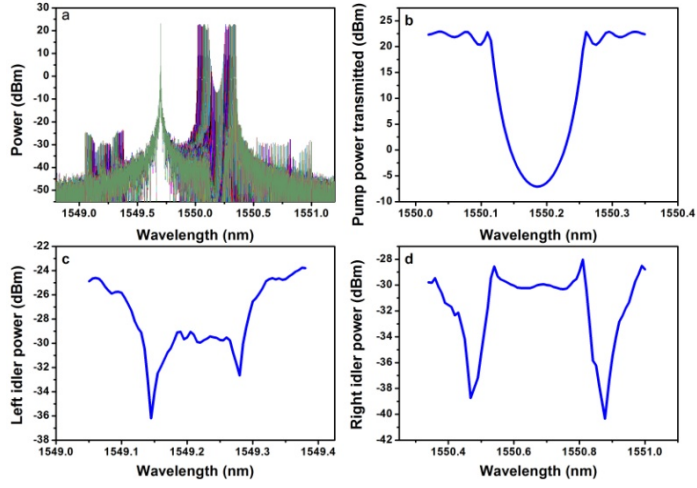

**Figure 3. Simulation results of degenerated FWM process with fast light.** Inset(a) shows the superposition of FWM spectrums with different pump wavelengths, and (b),(c),(d) shows the transmission spectrum of pump power, left and right idler spectra respectively.

**Optimizing the fast light effect.** According to the above derivation, the key parameter of FBG to generate the FWM in the bandgap is the dispersion coefficient to manipulate the fast/slow light effect. In Figure 4, the product of coupling coefficient $\kappa$ and FBG length $L^{28}$ shows a monotonous increasing tendency with advancement in the photonic bandgap and thus can represent the induced dispersion. By changing a series of FBG parameters, $\kappa L$ decreases with power of idler wave in both sides, either when fast light effect is considered or not considered. However, with more dispersion induced, the power difference between situations with and without fast light effect becomes larger, a limited maximum power difference of 20dB can be observed. However, due to the noise background of pump laser, a much larger idler wave power enhancement is convinced.

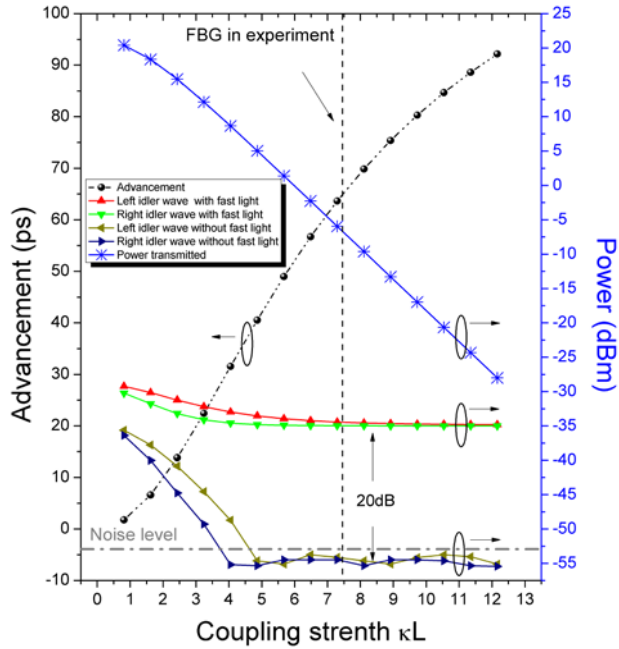

**Figure 4. Simulation result of FBG parameters.** $\kappa L$ shows an increasing tendency with advancement in the photonic bandgap and thus can represent the induced dispersion. Dash line shows the parameter of FBG in experimental configuration.

**Experimental configuration.** Figure 5 shows the experiment scheme. Two incident light with different wavelength were emitted by two tunable laser sources (TLS). Two Erbium-doped fiber amplifiers (EDFAs) amplify two lasers at the pump and signal waves of FWM. A polarization-independent isolator followed protects EDFAs from the reflected light by the FBG. The above special-designed FBG with a length of 10cm, Bragg wavelength at 1550.186nm, 3dB bandwidth 0.158nm provides the incident light aimed at its bandage an appreciable delay as well as advancement for those in the photonics bandgap. Through 1% output-port of an 99:1 optical coupler(OC), the ultimate FWM spectrum can be observed and analyzed by an optical spectrum analyzer (OSA).

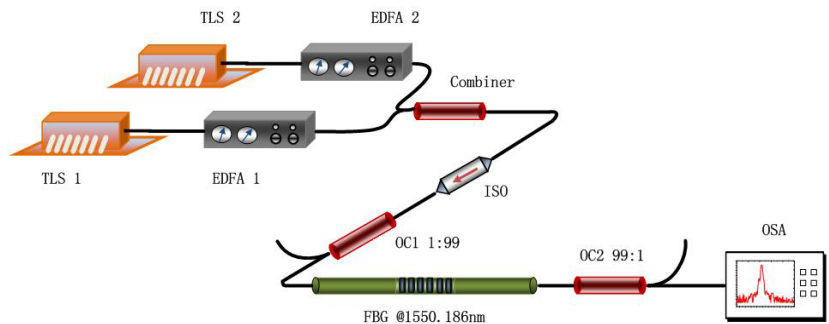

**Figure 5. Experimental configuration.** Scheme used to study the change of fiber FWM efficiency based on slow and fast light effect. TLS: Tunable Lasers. EDFA: Erbium-Doped Fiber Amplifier. ISO: Isolator. FBG: fiber Bragg grating. OC: Optical Coupler. OSA: Optical Spectrum Analyzer.

As the result shows in Figure 6, in either side of idler wave spectrum, there exists several peaks and hollows. When the pump light aims at the edge of FBG of 1550.11nm, and thus it experiences a group delay of 68.48ps. The idler wave powers in both sides suffer a decrease, 1.826dB in shorter wavelength and 1.204dB in longer part respectively. When the pump light aims at the edge of FBG in a longer wavelength around 1550.26nm, and it experiences a group delay of 62.75ps, the idler wave powers in both sides suffer a decrease of 2.023dB in shorter wavelength and 1.507dB in longer part respectively. Whereas, while the pump light is fully reflected and thus gets an advancement of 68.71ps in the photonic bandgap, the fast light effect compensates the power of idler wave in shorter wavelength partly and even helps the longer part back up to the same level outside the photonic bandgap. In contrast, due to the phase-mismatch and short length of fiber[34], the FWM generation was not observed in this experiment with the substitution of a single-mode fiber with the same length to the FBG. This experimental result shows a good agreement with the prediction in our simulation.

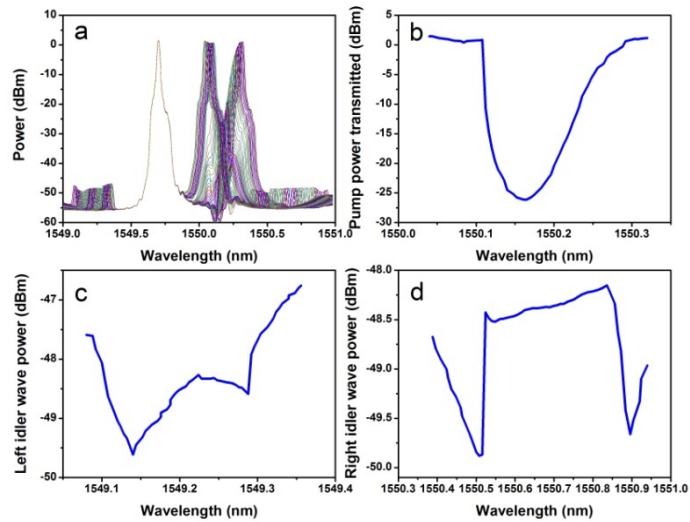

**Figure 6. Experimental results**, the change of idler wave power though the 1% output-port of an 99:1 OC owing to the slow/fast light effect in FWM. Inset(a) shows the superposition of FWM spectrums with different pump wavelengths, and (b),(c),(d) shows the transmission spectrum of pump power, left and right idler spectra respectively.

## Discussion

As is known to all, chromatic dispersion induced in the photonic bandgap of a typical FBG is roughly 6 orders of magnitude larger, which is of the same magnitude with the one we have tested, than that of bare fiber at 1550nm[35]. Slow/fast light effect is easy to observe in this strong dispersion regime, either normal or anomalous. The rapid variation of the phase and thus caused slow light group delay and FWM efficiency decrease are limited to a very small region at the band edge of FBG. In contrast, since the fast light effect together with the induced extra phase shift causes the satisfaction of phase-match, FWM process can be observed in the whole photonic bandgap. Nowadays FBG with a group delay of ~20ns at the edge has been reported[36] and it is now able to fabricate a meters-long-FBG, with a large index modulation, strong induced chromatic dispersion, long interaction length, several orders of magnitude FWM enhancement can be expected. Furthermore, although it is for the first time that we observe FWM process in the photonic bandgap, this theory also suits photonic bandgaps in other structures, like photonic crystal etc. We believe that, a series of other propagation-related nonlinear effects in the photonic bandgap, which were regarded as impossible, can be observed by means of slow/fast light effect.

**Conclusion**

In conclusion, we have proposed and demonstrate the FWM generation in the photonic bandgap and energy conversion of the standing waves to channels out of the photonic bandgap, which is traditionally thought to be forbidden. By introducing extra phase shift via fast light effect, we have made this impossible nonlinear process possible to occur in the photonic bandgap. The enhanced FWM has been observed both in simulation and experiment. With the help of the extra phase induced by fast light effect, not only FWM but also many other nonlinear effects can be expected in the photonic bandgap. Such a way may be used to detect optical signals that usually filtered out by a photonic bandgap filter.

**Methods**

**Transmittivity and group delay of fiber Bragg grating.** Gain/loss and group delay of the Gaussian-apodized fiber Bragg grating with a length of 10cm was measured by Agilent 86038B photonic dispersion and loss analyzer.

Optical phase as well as dispersion can also be derived simultaneously.

**Simulation results of four wave mixing spectra with and without fast light.** Simulation was conducted by *OptiSystem* with a configuration similar to the experimental one. Power of both TLSs were fixed at 200mW and pumped into a FBG with the Bragg wavelength of 1550.19nm and the bandwidth of 0.15nm. Signal wavelength is fixed at 1549.7nm and the pump wave sweeps from 1550.02nm to 1550.35nm to cover the photonic bandgap of FBG. With the dispersion of FBG considered or not considered, we superimpose the FWM spectrum in different pump wavelengths to form figure 2 and figure 3.

**Optimizing the fast light effect.** Simulation was conducted by *OptiGrating*. By changing the index modulation of FBG from $10^{-4}$ to $1.5 \times 10^{-3}$, a monotonous increasing tendency of advancement in the photonic bandgap with the product of coupling coefficient $\kappa$ and FBG length $L$ is derived. By changing a series of FBG with different index modulation in *OptiSystem*, relationship between $\kappa L$ and power of idler wave in both sides can be achieved, either when fast light effect is considered or not considered.

**Experimental configuration.** According to the scheme in Figure 5, experiment was carried out by setting the wavelength of one input light from TLS at 1549.7nm, which was not reflected by FBG. The wavelength of the other input light is fine tuned from 1550.04nm to 1550.32nm, which covers the whole range of the bandgap region of the FBG. The spectrum of the light from output port of OC1 is also monitored by OSA to ensure that two incident lights are amplified to almost same intensity and their power fluctuations during the experiment are ignorable. Through the spectrum of output port of OC2, changes of the idler wave power during the sweep is observed.

**Acknowledgements**

The authors acknowledge the support from the National Natural Science Foundation of China (Grants 61178014/11274231/61308003).

**Author contributions**

Cheng Feng performed experiment, simulation and analysis. Li Zhan supervised the project. The manuscript was prepared by Cheng Feng, Hao Luo and Li Zhan. All authors discussed the results and substantially contributed to the manuscript.

**Competing financial interests**

The authors declare no competing financial interests.

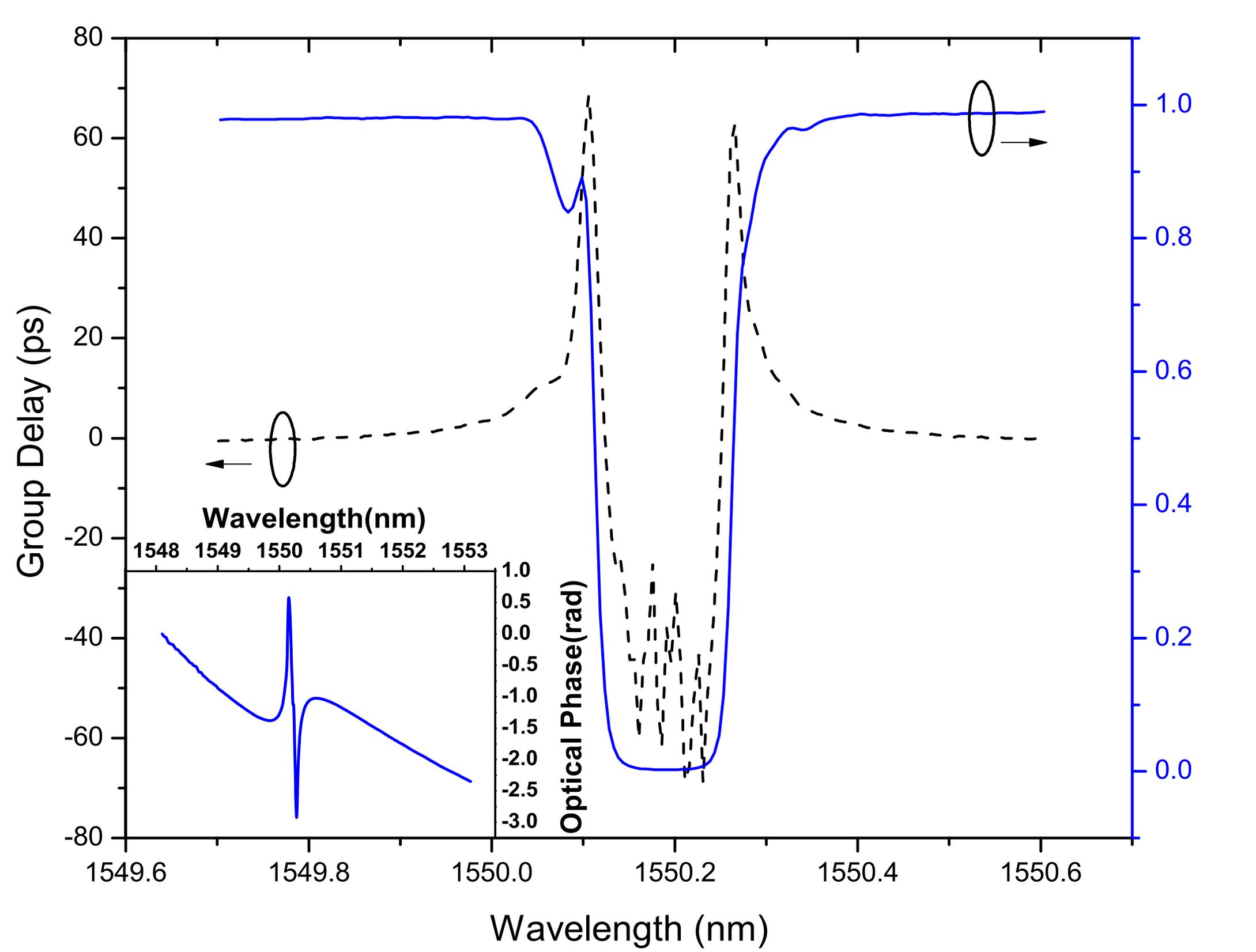

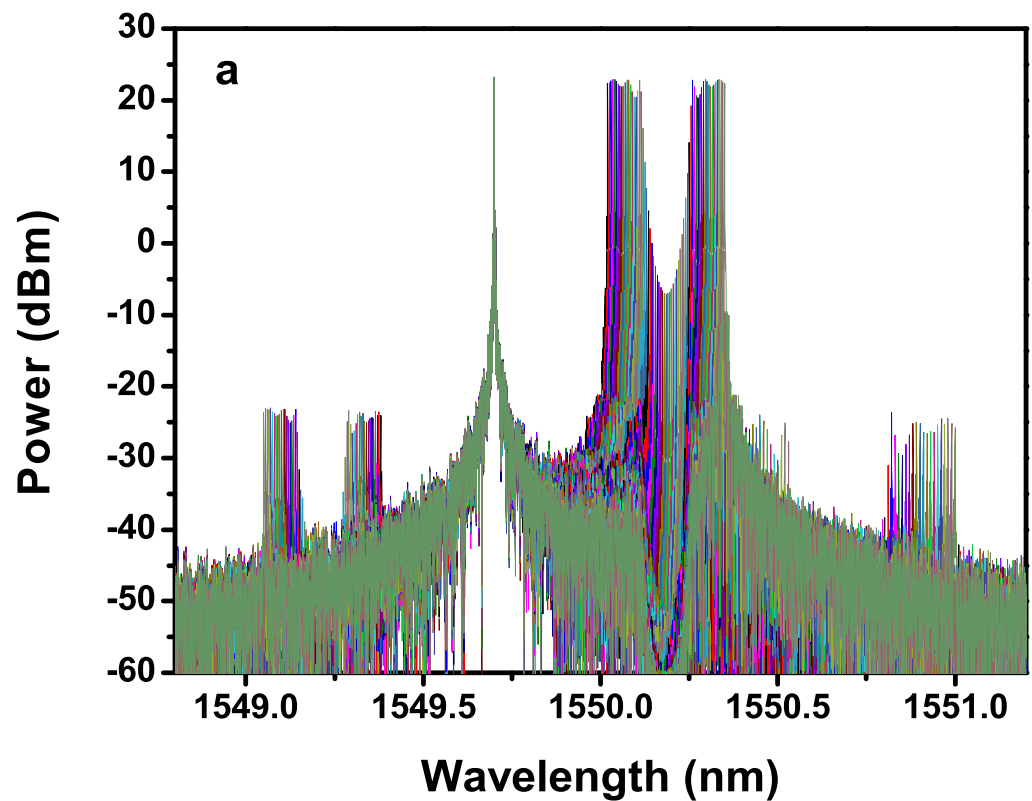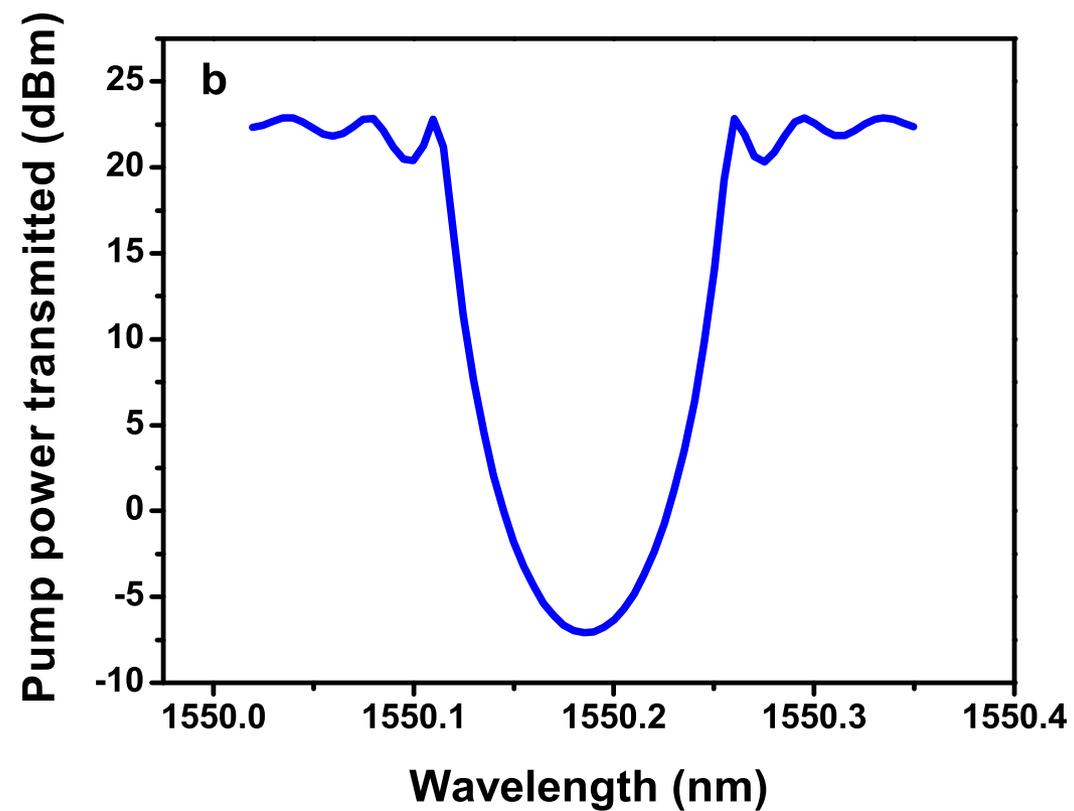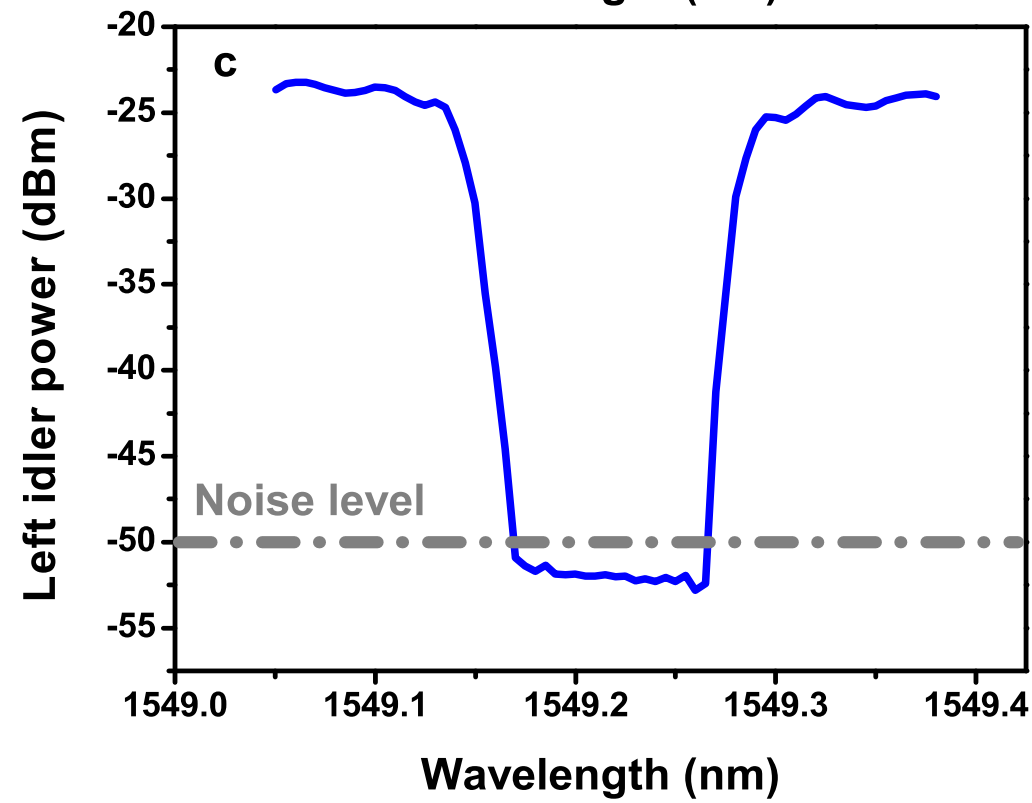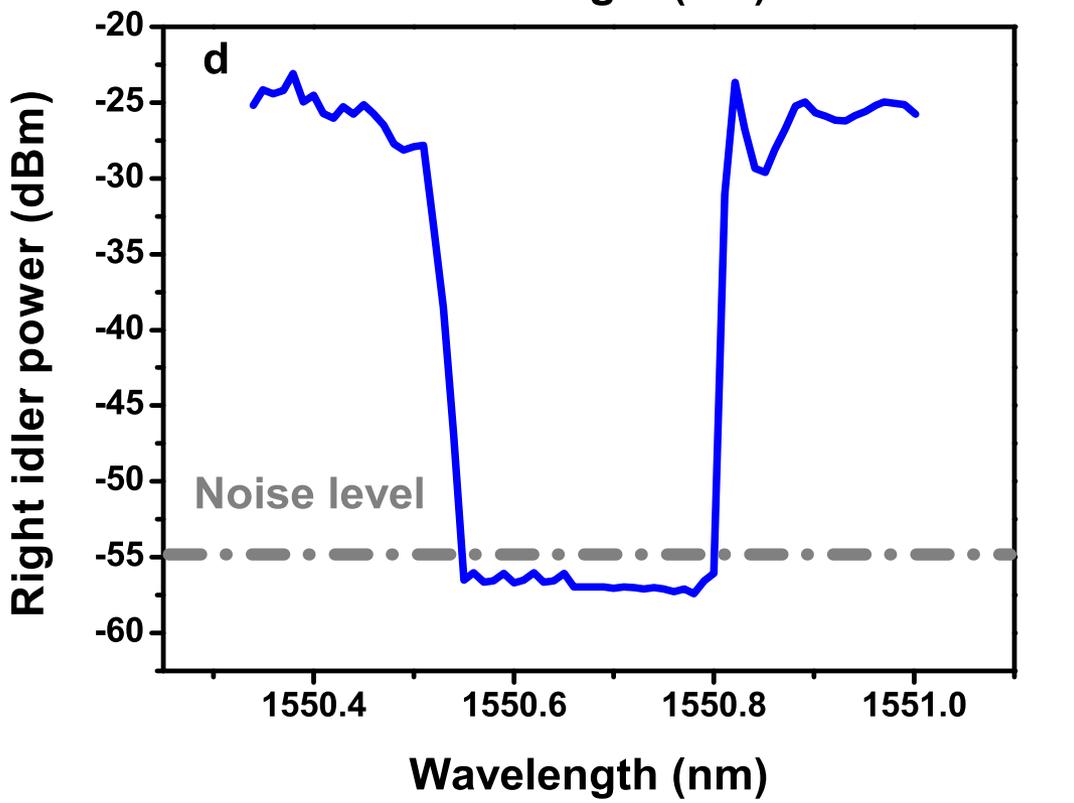

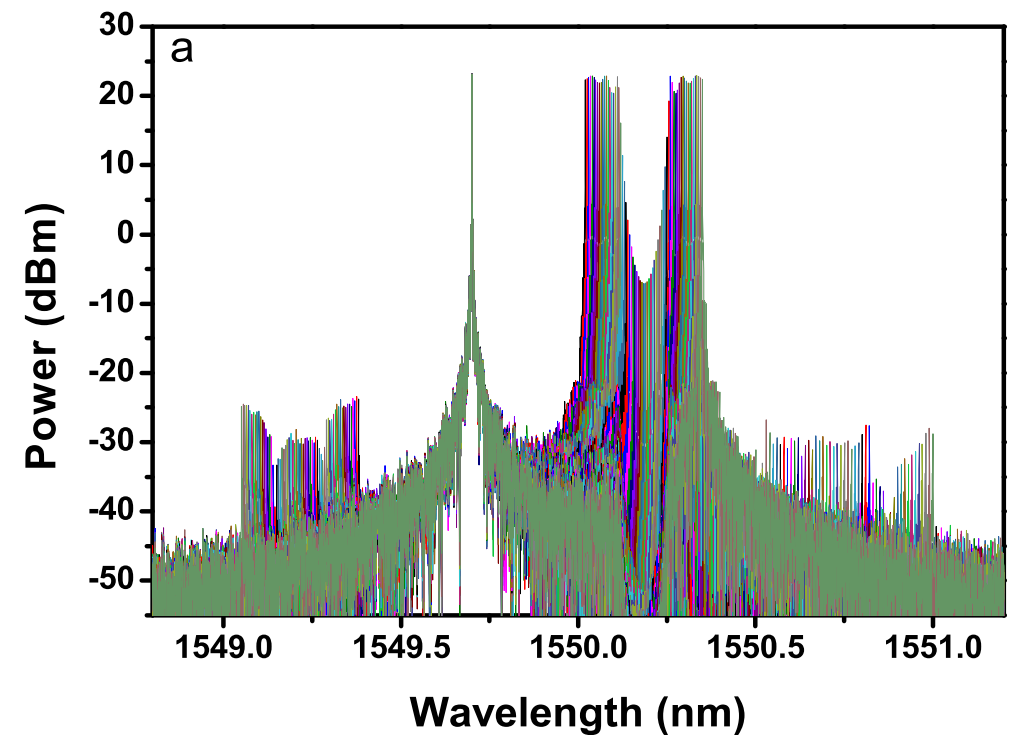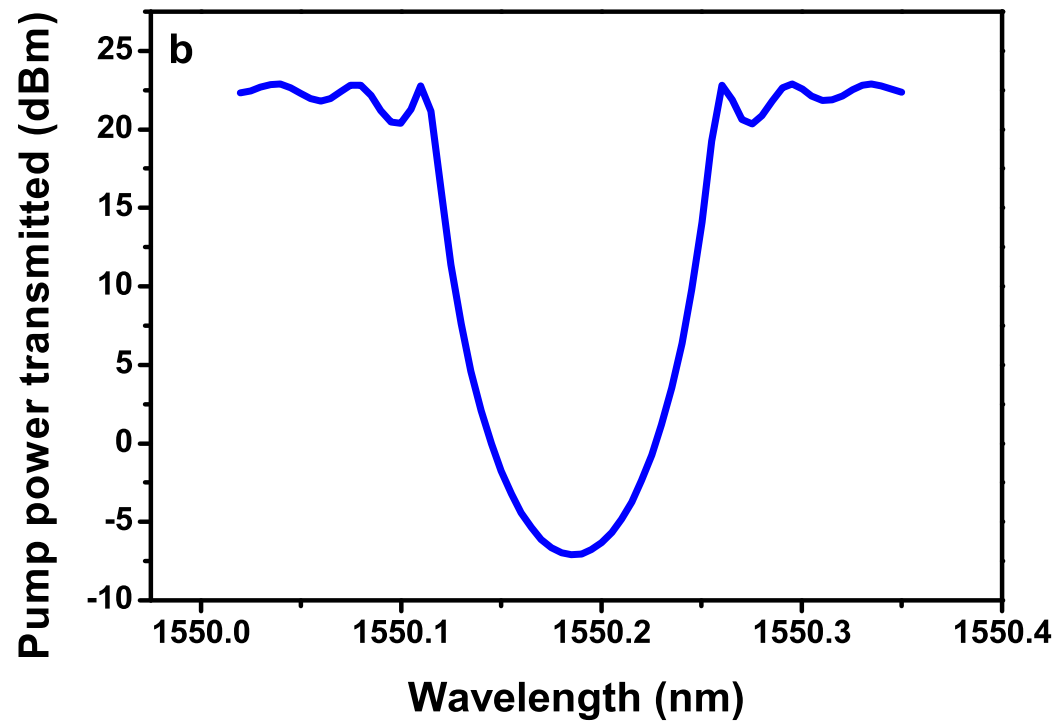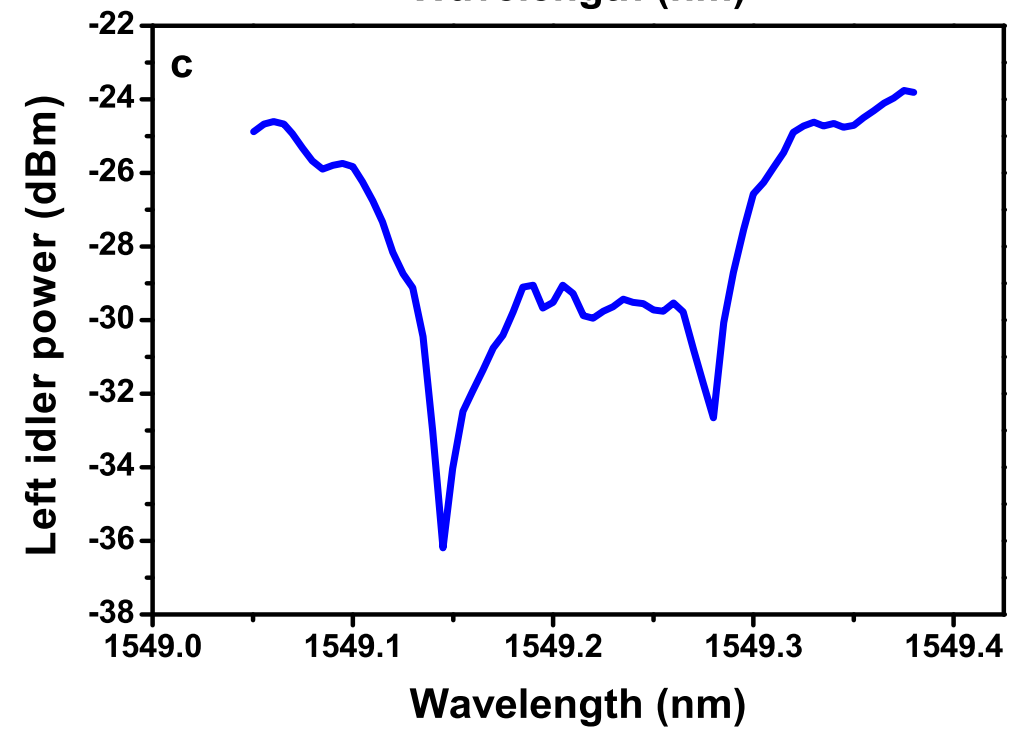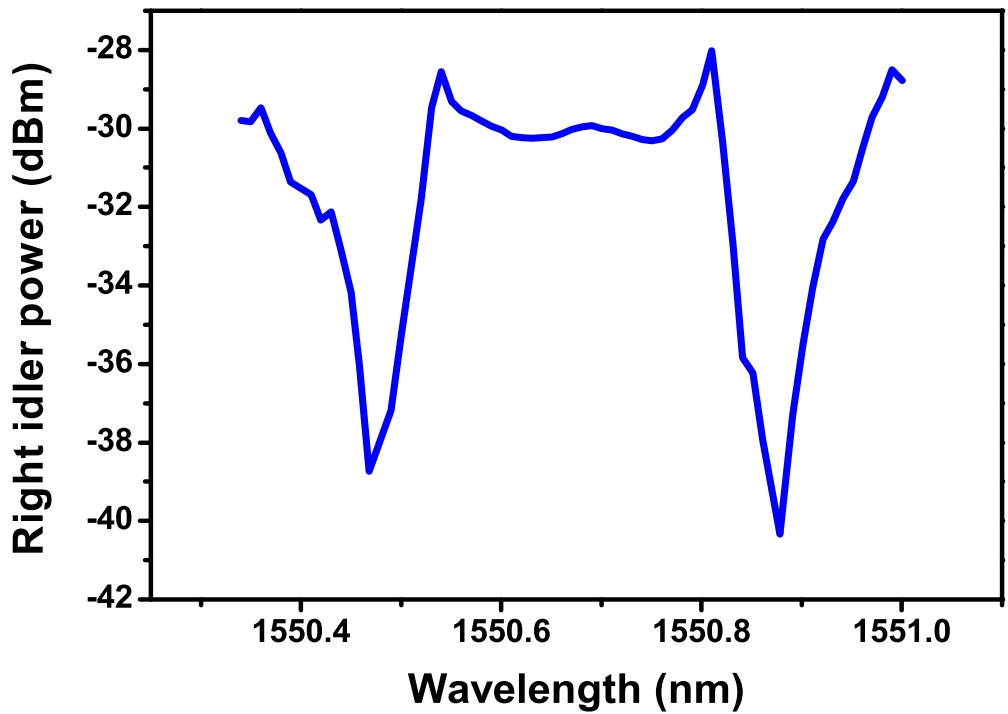

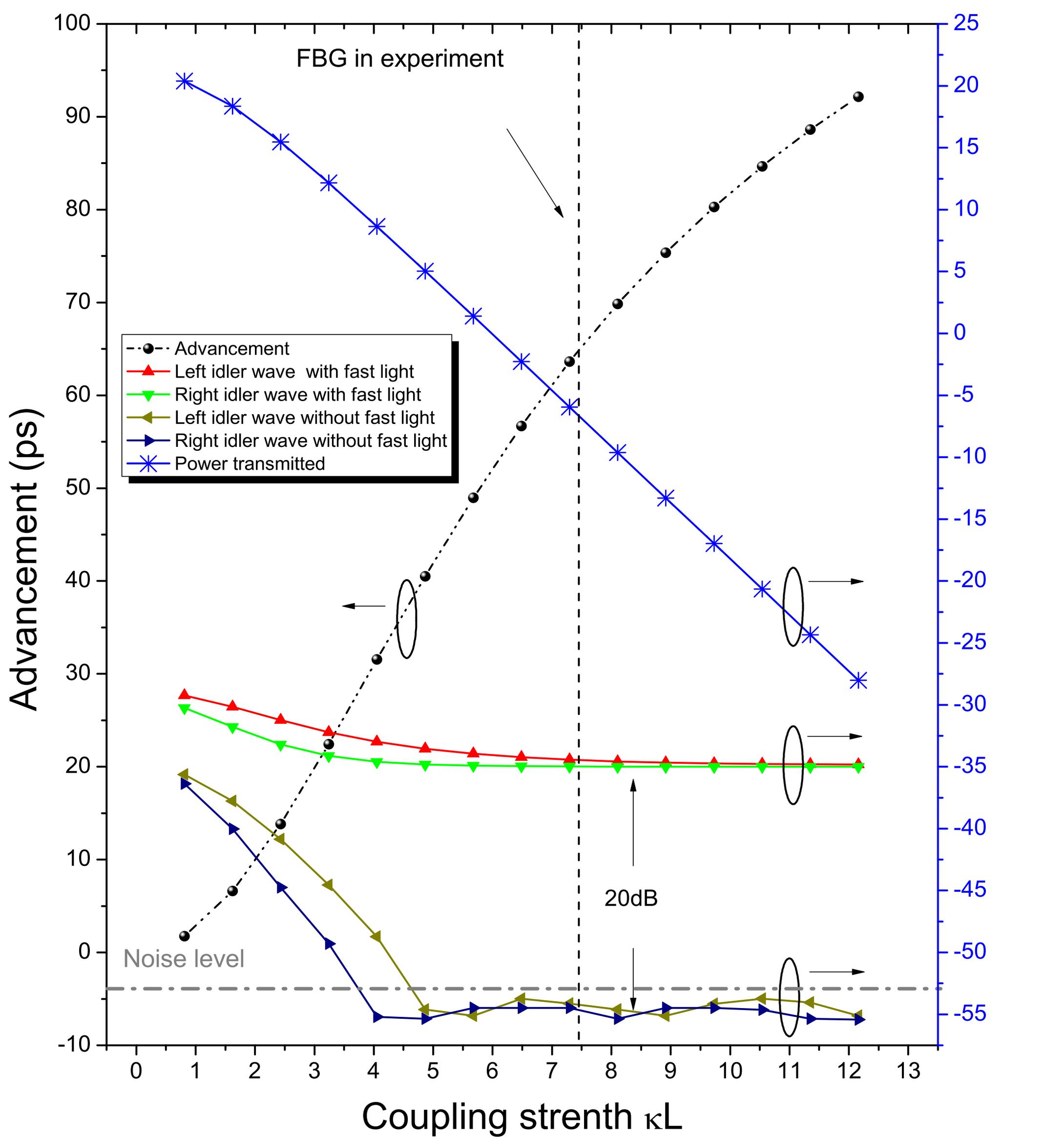

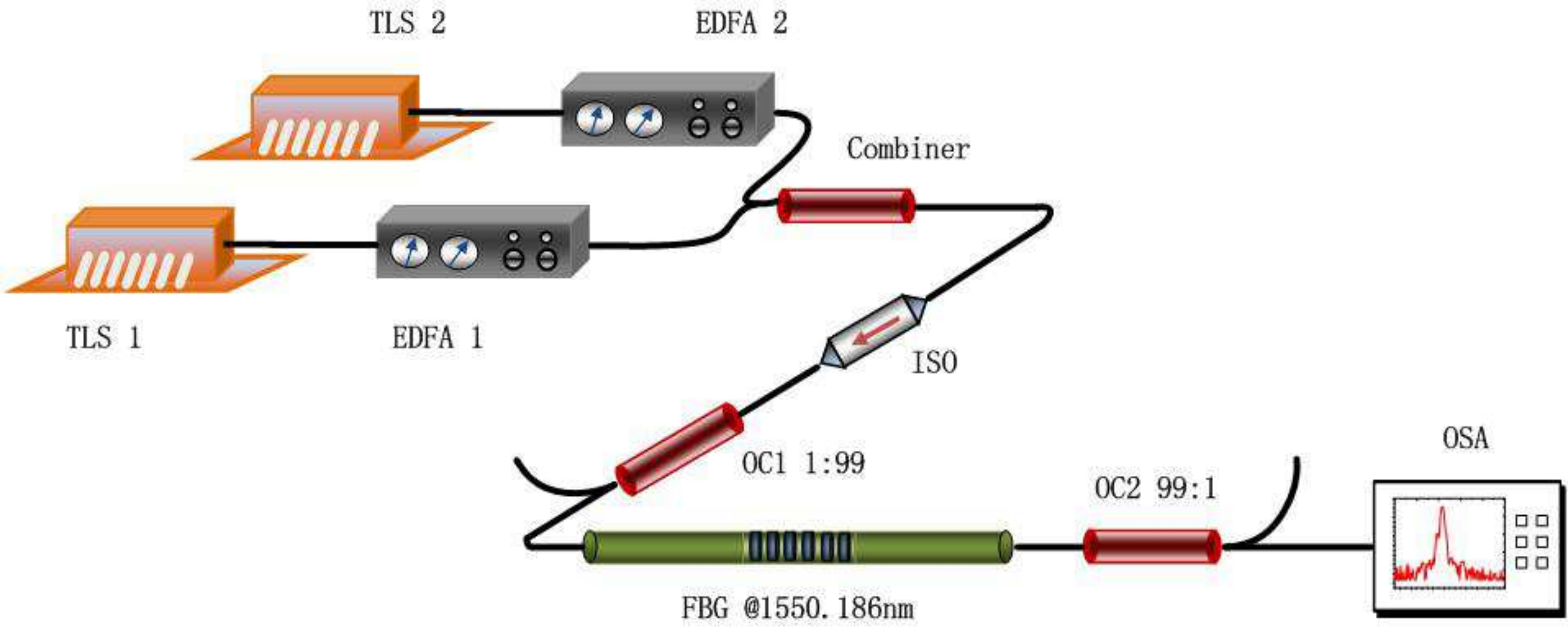

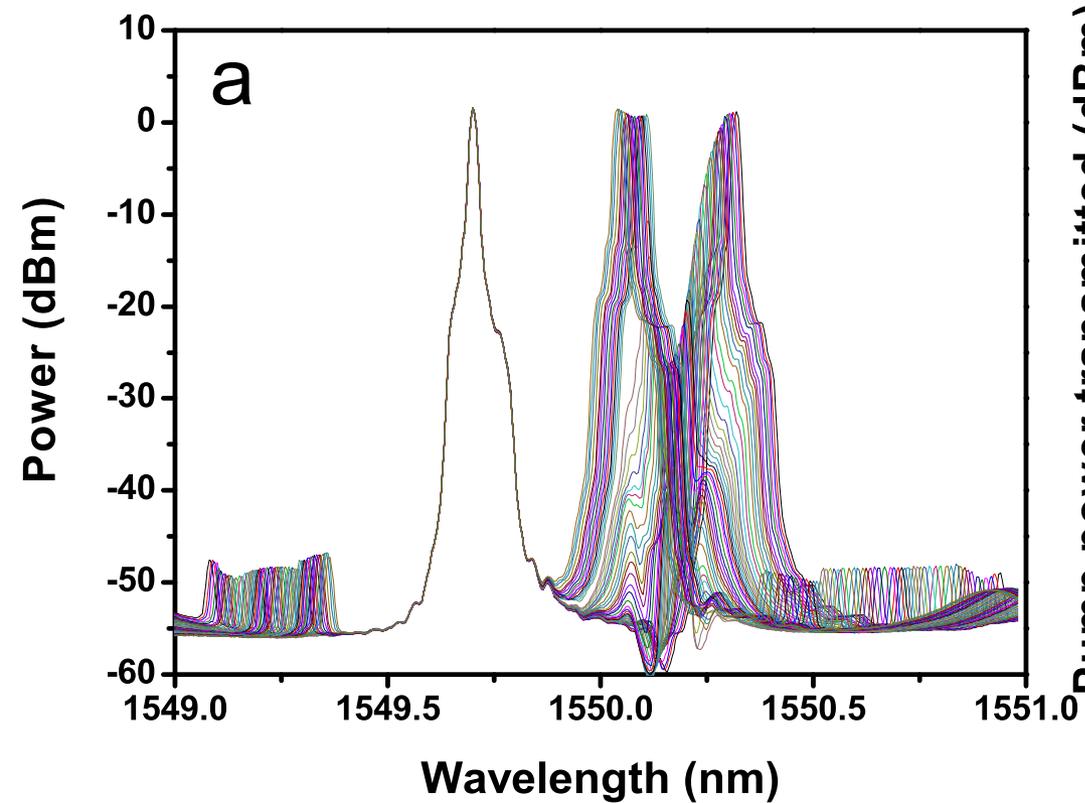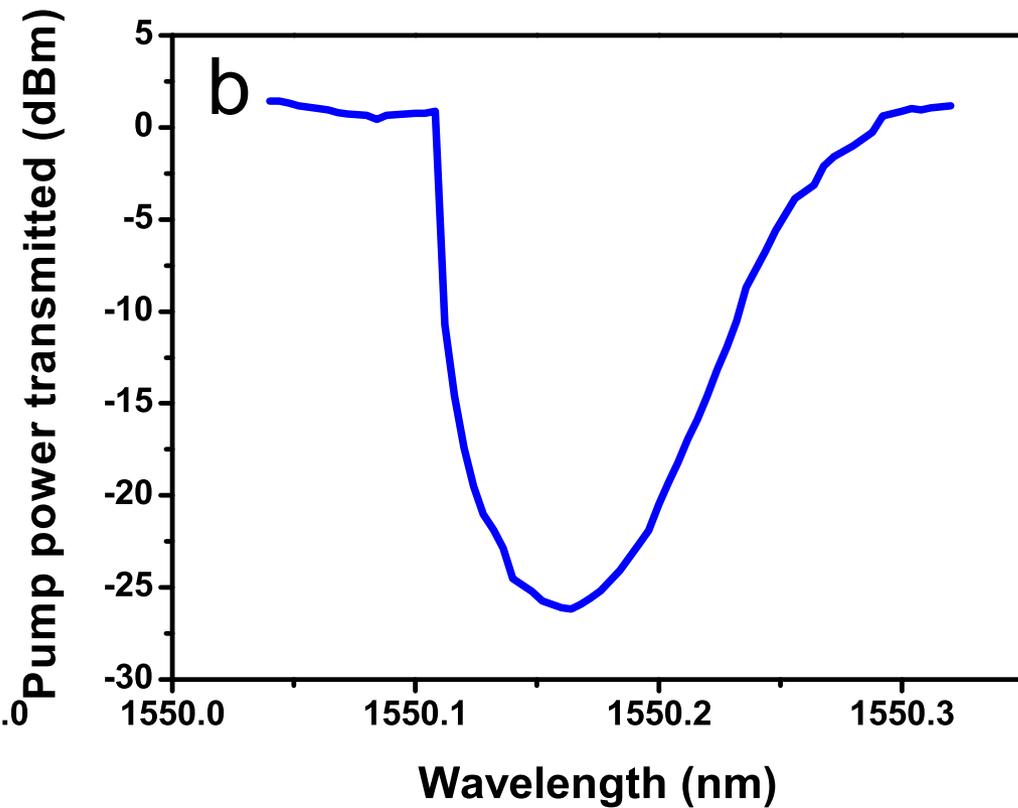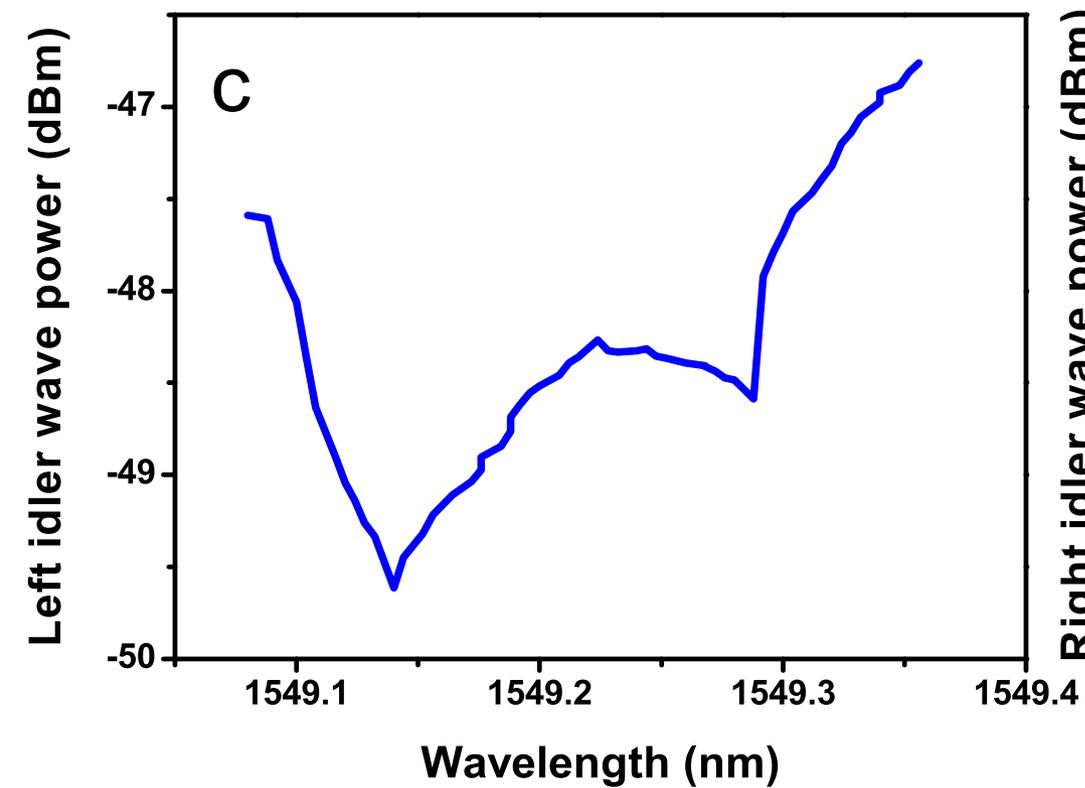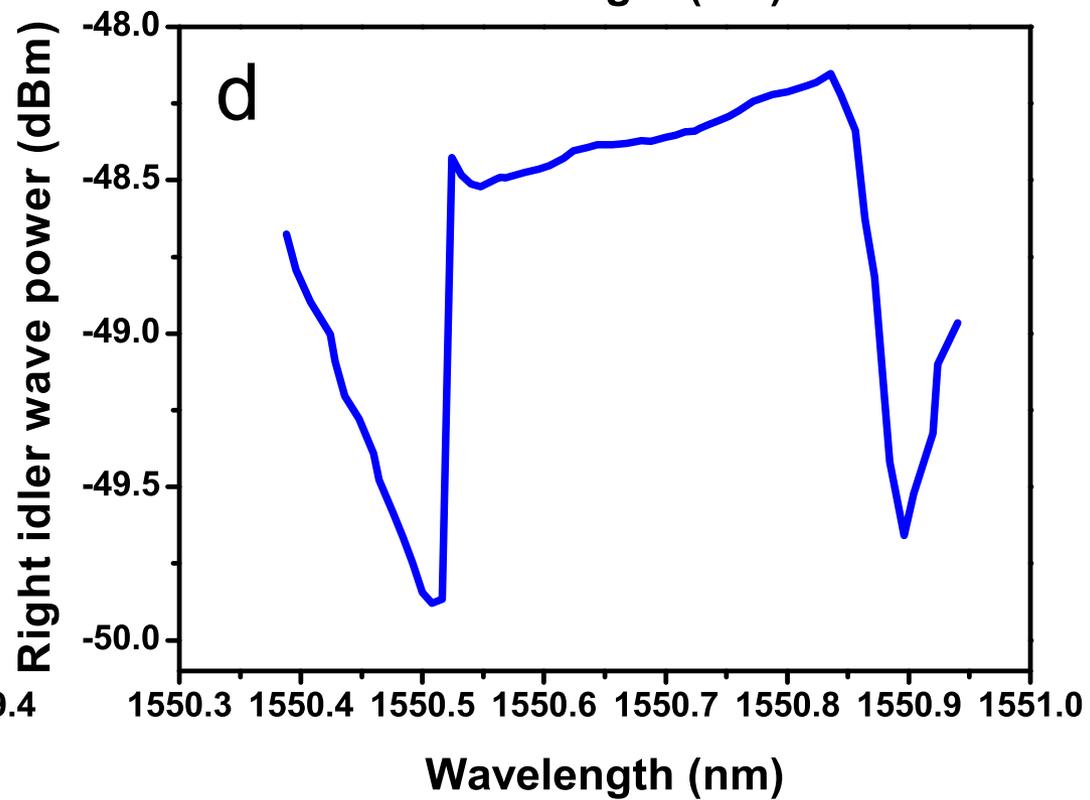